\documentclass[sigconf]{acmart}
\usepackage{booktabs}
\usepackage{longtable}
\AtBeginDocument{%
  }

\setcopyright{acmlicensed}
\copyrightyear{2025}
\acmYear{2025}
\begin{document}

\title{State of the Quantum Software Engineering Ecosystem
}


\author{Nazanin Siavash}
\email{nsiavash@uccs.edu}
\orcid{0009-0000-4177-0632}
\affiliation{
  \institution{University of Colorado Colorado Springs (UCCS)}
  \country{United States}
}

\author{Armin Moin}
\email{amoin@uccs.edu}
\orcid{0000-0002-8484-7836}
\affiliation{
  \institution{University of Colorado Colorado Springs (UCCS)}
  \country{United States}
}

\renewcommand{\shortauthors}{Siavash and Moin}

\begin{abstract}
We study the current state of the Quantum Software Engineering (QSE) ecosystem, focusing on the achievements, activities, and engagements from academia and industry, with a special focus on successful entrepreneurial endeavors in this arena. Our research methodology is a novel one, featuring the state-of-the-art in Artificial Intelligence (AI), namely Large Language Models (LLMs), especially Generative Pretrained Transformers (GPT). We use one of such models, namely the OpenAI GPT-5 model, through the ChatGPT tool. The goal is to identify institutions and companies that are highly active and have achieved distinguished results in QSE, evidenced by peer-reviewed publications or raised capital in the venture capital market.  
\end{abstract}



\keywords{quantum computing, quantum software engineering, quantum software ecosystem, quantum software startups, gpt-5, chatgpt}


\maketitle

\section{Introduction} \label{sec:introduction}

Murillo et al. \cite{Murillo+2025} identified several key areas of Quantum Software Engineering (QSE) and determined the most relevant open challenges that should be addressed in the next years. In this work, we focus on identifying active institutions and companies that have been engaged within the QSE ecosystem to enable synergies and future collaborations that we hope will help address the challenges in the coming years. 


As described in Section \ref{sec:methodology}, we take a fundamentally different approach from the existing work, such as the work of Ruane et al. \cite{Ruane+2025}. First, we focus only on QSE. Second, we deploy an innovative research method and tool, based on cutting-edge AI tools enabled by Large Language Models (LLMs). Additionally, instead of relying on patents, we measure the success of entrepreneurial activities based on the trust of investors, measured by the amount of capital raised by them.

The rest of this paper is structured as follows. Section \ref{sec:related-work} reviews related work in this area, whereas Section \ref{sec:methodology} elaborates on our research methodology. In Section \ref{sec:results}, we present the research outcomes of this study. Finally, we conclude and propose future work in Section \ref{sec:conclusion-future-work}.


\section{Related Work} \label{sec:related-work}
The Quantum Index Report (QIR) 2025 from the Massachusetts Institute of Technology (MIT) \cite{Ruane+2025} reported on the state of academic research, patents, venture funding, corporate communication, policy, workforce, education, public opinion, etc., regarding quantum technologies around the globe. The report highlighted that we are on the verge of the second quantum revolution. While the first quantum revolution has enabled groundbreaking technologies, such as semiconductors, lasers, MRI machines, and atomic clocks, the second revolution is expected to focus on controlling and engineering quantum systems directly. For instance, the ability to program with qubits for computation and information processing, leveraging their unique and unprecedented characteristics — such as superposition, entanglement, and interference —will be highly disruptive for a handful of algorithms, some of which are widely deployed in various applications. The engineering discipline that is concerned with this new area is QSE. However, it also inherently involves classical software engineering (SE), as well as principles and practices for the engineering of hybrid quantum-classical systems \cite{Sepulveda+2024}.

\section{Methodology} \label{sec:methodology}
First, we consider author affiliations from the latest edition/volume of conference proceedings and journal articles from the following international, accredited, and peer-reviewed scientific publication venues: (i) ACM Transactions on Quantum Computing with a Google Scholar h-index of 33. (ii) The IEEE Quantum Computing and Engineering (QCE) conference (also known as IEEE Quantum Week) with a Google Scholar h-index of 36. (iii) Quantum Machine Intelligence with a Google Scholar h-index of 36. (iv) Quantum Information Processing with a Google Scholar h-index of 47. (v) IEEE Transactions on Quantum Engineering with a Google Scholar h-index of 48. (vi) npj Quantum Information with a Google Scholar h-index of 73.

We come up with the list above after searching with the keyword \textit{quantum} on Google Scholar Metrics and filtering out venues based on the following inclusion and exclusion criteria: i) Include all venues that contain \textit{computing}, \textit{software}, \textit{information}, \textit{machine}, or \textit{engineering} in their names. ii) Exclude all venues that have a Google Scholar h-index of below 30 or do not include any of the stated keywords in their names. Finally, due to the unique position of the International Conference on Software Engineering (ICSE) in the SE research community, as the top-ranked international conference on SE, and the existence of its Q-SE workshop, we decided to append the ICSE Q-SE workshop to the list above. We then use OpenAI's ChatGPT with the GPT-5 model with appropriate prompt engineering to extract the desired information. We explain the details in Section \ref{sec:results} below. We randomly verify a number of cases to make sure the LLM-based tool does not \lq{}hallucinate\rq{} or, when that happens, capture and report it.

Furthermore, we acknowledge that, for startups, peer-reviewed publications may not be relevant. Therefore, we skip looking at the research publications. Instead, we directly ask OpenAI's ChatGPT with the GPT-5 model to extract a list of top quantum software-based startups worldwide, which have raised at least \$1M of funds. This approach is different from the typical patent-oriented approach, for example, as in Ruane et al. \cite{Ruane+2025}. Note that not all software innovations are patentable. Moreover, the quantity of patent activity does not necessarily translate to a higher impact or value in the quantum software ecosystem. For instance, many open-source artifacts, libraries, and tools have a significant impact without being patented.


\section{Research Results} \label{sec:results}

Tables \ref{tab:academia1}, \ref{tab:academia2}, \ref{tab:academia3}, \ref{tab:academia4}, and \ref{tab:academia5} present the results of our study for Institutions of Higher Education (IHE), also known as Universities (Univ.), as well as academic research centers. This excludes government labs, such as federally funded labs/research institutions in the U.S. We provided the GPT-5 LLM with the following prompt series through the ChatGPT tool's User Interface (UI) and then manually reviewed and combined the obtained outcomes: (i) \textit{Take a look at the latest volume of \$journal\$, focused on quantum software, and list all types of papers (e.g., technical, and survey papers) authored by institutions of higher education and academic research centers.} (ii) \textit{Take a look at the latest edition of \$conference\$, focused on quantum software, and list all types of papers (e.g., technical, posters, workshops, and survey papers) authored by institutions of higher education and academic research centers.} In prompt (i), \$journal\$ $\in$ \{ACM Transactions on Quantum Computing, Quantum Machine Intelligence, Quantum Information Processing, IEEE Transactions on Quantum Engineering, npj Quantum Information\}. Moreover, in prompt (ii), \$conference\$ $\in$ \{IEEE Quantum Computing and Engineering (QCE) conference (also known as IEEE Quantum Week), ICSE Q-SE workshop\}. Since complete author affiliations were not available for all papers of QCE, we manually reviewed each paper to identify the academic affiliations.

In Table \ref{tab:industry}, we see the industrial participants in this ecosystem. This list has been achieved using a similar method as explained above. However, we augmented the corresponding prompts as follows: \textit{...and limit it to articles authored by at least one person affiliated with industrial research centers or companies.}

Furthermore, since publishing in peer-reviewed venues may not always be a priority for industrial participants in the QSE ecosystem and also due to the fact that some of them may be missing in the latest editions of the conferences and the most recent volumes of the journals, we decided to explicitly ask ChatGPT the following: \textit{Create a table listing quantum software-focused industrial companies worldwide. The table should include the following columns: company name, artifact (e.g., library, tool, sdk), country, state or province, and official website.} For the sake of space, we summarize the outcomes of this step and only name the companies and their artifacts: (i) IBM (Qiskit), (ii) Microsoft (Azure Quantum QDK/Q\#), (iii) Google (Cirq), (iv) Rigetti Computing (pyQuil and QCS SDK), (v) Quantinuum (tket compiler/SDK), (vi) QC Ware (Forge platform), (vii) Zapata AI (Orquestra platform), (viii) Strangeworks (Strangeworks Platform), (ix) Q-CTRL (Boulder Opal), (x) Xanadu (PennyLane SDK), (xi) Agnostiq (Covalent workflow orchestrator), (xii) 1QBit (1Qloud/QEMIST software suite), (xiii) Riverlane (Deltaflow.OS/quantum OS), (xiv) HQS Quantum Simulations (HQS software/HQStage), (xv) PASQAL (Pulser), (xvi) QuEra (Bloqade/Bloqade.jl), (xvii) QunaSys (QURI Parts), (xviii) Terra Quantum (TQ42 Studio), (ixx) Quantum Computing Inc. (Qatalyst).

Last but not least, following the method explained in Section \ref{sec:methodology} for startups, the following list was generated (which in part overlaps with the above-mentioned one): (i) SandboxAQ (CA, US), (ii) Multiverse Computing (Spain), (iii) Classiq (Israel), (iv) Q-CTRL (Australia), (v) Riverlane (UK), (vi) Zapata AI (MA, US), (vii) Phasecraft (UK), (viii) QC Ware (CA, US), (ix) Strangeworks (TX, US), (x) Algorithmiq (Finland), (xi) BlueQubit (CA, US), (xii) Agnostiq (Canada), and (xiii) Qruise (Germany).


\section{Conclusion and Future Work} \label{sec:conclusion-future-work}
We have studied the state of the QSE ecosystem, particularly the engagement of academia, industry, and entrepreneurs. One particular challenge with the LLM-based tool, ChatGPT, was with respect to the reproducibility of the results. For instance, for startups, we obtained several variants of the list each time we executed the same prompt. Furthermore, in some cases, multiple authors shared the same name. These cases were reviewed manually.

All in all, we have found out that similar to the broader area of Quantum Computing (QC), the emerging new subarea of QSE is growing at a fast pace but has not yet been established in various arenas in academia. For example, on Cornell University's pre-print repository, \textit{arXiv}, there is no QC category present yet, let alone QSE. Such papers are still categorized under Quantum Physics. Similarly, on the \textit{csrankings.org} website used to rank Computer Science departments across the globe, all the mentioned venues in Section \ref{sec:methodology} except for ICSE are absent in their publication criteria.


\begin{acks}
We used generative AI models and tools, including OpenAI’s GPT-5 model through ChatGPT, to assist with generating and revising content. This work is funded by a grant from the Colorado OEDIT.
\end{acks}

\bibliographystyle{ACM-Reference-Format}
\bibliography{refs.bib}

\onecolumn
\begin{table}[htbp!]
\centering
\caption{Engagement from academia - Part I}
\label{tab:academia1}
\begin{tabular}{|p{4cm}|l|p{1cm}|p{2.5cm}|p{8cm}|}
\hline
\textbf{Univ./research centers}&\textbf{Count}& \textbf{Country} & \textbf{Province / State} & \textbf{Collaborators }\\ \hline
University of Saskatchewan&3& Canada& Saskatchewan& University of Prince Edward Island\\ \hline
University of Prince Edward Island&2& Canada& Prince Edward Island& University of Saskatchewan\\ \hline
 Fraunhofer Institute for Industrial Mathematics (ITWM)& 2& Germany& Kaiserslautern&Fraunhofer IAIS; TU Dortmund, \\\hline
 Fraunhofer IAIS& 1& Germany& Sankt Augustin&Fraunhofer ITWM; TU Dortmund\\\hline
 TU Dortmund University& 1& Germany& Dortmund&Fraunhofer ITWM; Fraunhofer IAIS\\\hline
 Vellore Institute of Technology (School of Computer Science and Engineering)& 1& India& Tamil Nadu&None\\\hline
 Artificial Intelligence Research Institute (IIIA-CSIC)& 1& Spain& Catalonia (Barcelona)&Universitat Autònoma de Barcelona\\\hline
 Universitat Autònoma de Barcelona& 1& Spain& Catalonia (Barcelona)&Artificial Intelligence Research Institute (IIIA-CSIC)\\\hline
 University of California – Berkeley& 3& USA& California&Lawrence Berkeley National Laboratory, Argonne National Laboratory\\\hline
 University of Montpellier& 1& France& Montpellier&CERFACS, LIRMM, CNRS\\\hline
 University of Tennessee, Knoxville (UTK)& 1& USA& Tennessee&Oak Ridge National Laboratory\\\hline
Guangxi University of Science and Technology — School of Computer Science and Technology& 1& China& Guangxi&Jinggangshan University\\\hline
Jinggangshan University — School of Electronics and Information Engineering& 1& China& Jiangxi&Guangxi University of Science and Technology \\\hline
Boise State University — Department of Computer Science& 1& USA& Idaho&None\\\hline
Hanyang University — Graduate School, Dept. of Electrical and Electronic Engineering (Seoul)& 1& South Korea& Seoul (Special City)&Hanyang University ERICA (Ansan)\\\hline
Hanyang University ERICA — School of Electrical Engineering (Ansan)& 1& South Korea& Gyeonggi-do&Hanyang University (Seoul)\\\hline
Westfälische Hochschule (University of Applied Sciences), Bocholt — Dept. of Business \& Information Technology& 1& Germany& North Rhine–Westphalia&Oak Ridge National Laboratory\\\hline
Anhui University — School of Integrated Circuits& 1& China& Anhui&Anhui Normal University\\\hline
Anhui Normal University — School of Computer and Information& 1& China& Anhui&Anhui University — School of Integrated Circuits\\\hline
 Pennsylvania State University& 2& USA& Pennsylvania&Quantum Economic Development Consortium, SRI International, Vanderbilt University, Nashville, Coherent Computing Inc., Sandia National Laboratories, Research Institute of Advanced Computer Science, Universities Space Research Association, Purdue University,  Los Alamos National Laboratory, Quantum Circuits Inc, Quantum Economic Development Consortium Technical Advisory Committee on Standards and Performance Metrics, Lawrence Berkeley National Laboratory, University of Central Florida, Coherent Computing Inc., Purdue University\\\hline

\end{tabular}
\end{table}

\begin{table}[htbp!]
\centering
\caption{Engagement from academia - Part II}
\label{tab:academia2}
\begin{tabular}{|p{4cm}|l|p{1cm}|p{2.5cm}|p{8cm}|}
\hline
\textbf{Univ./research centers}&\textbf{Count}& \textbf{Country} & \textbf{Province / State} & \textbf{Collaborators }\\ \hline
 Vanderbilt University& 1& USA& Tennessee&Pennsylvania State University, Quantum Economic Development Consortium, SRI International,  Nashville, Coherent Computing Inc., Sandia National Laboratories, Research Institute of Advanced Computer Science, Universities Space Research Association, Purdue University,  Los Alamos National Laboratory, Quantum Circuits Inc, Quantum Economic Development Consortium Technical Advisory Committee on Standards and Performance Metrics\\\hline
 Purdue University& 2& USA& Indiana&Pennsylvania State University, Quantum Economic Development Consortium, SRI International,  Nashville, Coherent Computing Inc., Sandia National Laboratories, Research Institute of Advanced Computer Science, Universities Space Research Association, Purdue University,  Los Alamos National Laboratory, Quantum Circuits Inc, Quantum Economic Development Consortium Technical Advisory Committee on Standards and Performance Metrics, Vanderbilt University, Lawrence Berkeley National Laboratory, University of Central Florida, Coherent Computing Inc.,  Pennsylvania State University\\
 National Taiwan University& 3& Taiwan& Taipei&National Center for Theoretical Sciences,  Hon Hai Research Institute Columbia University, National Chiao Tung University, National Tsing Hua University\\\hline
 Keio University& 1& Japan& Yokohama&RIKEN Center for Computational Science, Kyushu University,   University College London, Kyoto University, NTT Corporation\\\hline
 Kyushu University& 4& Japan& Fukuoka&RIKEN Center for Computational Science, Keio University, Shanghai Maritime University, Stanford University, Kyushu Sangyo University, Nagoya University, \\\hline
 University of Innsbruck& 1& Austria& Innsbruck&Data:Lab, Volkswagen AG, ParityQC\\\hline
 Massachusetts Institute of Technology (MIT) — Department of Physics& 1& USA& Massachusetts&IBM Quantum, MIT-IBM Watson AI Lab\\\hline
 East China University of Science and Technology — Dept.\ of Computer Science and Engineering& 1& China& Shanghai&Shanghai Key Lab of Computer Software Testing \& Evaluating; Shanghai Software Center; University of Southern California\\\hline
 Shanghai Key Laboratory of Computer Software Testing \& Evaluating& 1& China& Shanghai&East China University of Science and Technology; Shanghai Software Center; University of Southern California\\\hline
 Shanghai Software Center& 1& China& Shanghai&East China University of Science and Technology; Shanghai Key Laboratory of Computer Software Testing \& Evaluating; University of Southern California\\\hline
 Karlsruhe Institute of Technology (KIT)& 1& Germany& Karlsruhe&None\\\hline
 Politecnico di Milano& 2& Italy& Milano&None\\\hline
 Kyushu Sangyo University& 1& Japan& Fukuoka Prefecture&Kyushu University\\\hline
 University of Stuttgart — Institute of Architecture of Application Systems (IAAS)& 1& Germany& Baden-Württemberg&University of Stuttgart (various departments)\\\hline
 William \& Mary & 1& USA& Virginia&None\\\hline
 University of British Columbia & 2& Canada& British Columbia &Quantum Science \& Exploratory Research, University of Naples, Los Alamos National Laboratory,  University of Jyväskylä, University of Illinois Urbana-Champaign\\\hline
 Fraunhofer Institute for Cognitive Systems IKS& 1& Germany& Munich&None\\\hline
 
\end{tabular}
\end{table}

\begin{table}[htbp!]
\centering
\caption{Engagement from academia - Part III}
\label{tab:academia3}
\begin{tabular}{|p{4cm}|l|p{1cm}|p{2.5cm}|p{8cm}|}
\hline
\textbf{Univ./research centers}&\textbf{Count}& \textbf{Country} & \textbf{Province / State} & \textbf{Collaborators }\\ \hline
Duke University& 4& USA& North Carolina&Georgia Tech, Univerity of Texas at Austin\\\hline
 Stanford University& 1& USA& California&Kyushu University, Shanghai Maritime University\\\hline
 Shanghai Maritime University& 1& China& Shanghai&Kyushu University, Standford University\\\hline
 Technical University of Munich& 5& Germany& Bavaria&Software Competence Center Hagenberg GmbH (SCCH), Munich Quantum Software Company GmbH, Yale University\\\hline
 Carnegie Mellon University& 2& USA& Pennsylvania&Yale University, QuEra Computing Inc, Amazon Web Services, University of California, Los Angeles\\\hline
 Yale University& 5& USA& Connecticut&Carnegie Mellon University, University of Chicago, Northwestern University, Technical University of Munich, 3Munich Quantum Software Company GmbH, Software Competence Center Hagenberg GmbH\\\hline
 Univerity of Texas at Austin& 2& USA& Texas&Duke University, Georgia Tech, University of Chicago\\\hline
 Georgia Tech&1 & USA& Georgia&Duke University, Univerity of Texas at Austin\\\hline
 University of California, Los Angeles& 1& USA& California&Carnegie Mellon University, QuEra Computing Inc, Amazon Web Services\\\hline
  EPFL&1 & Switzerland& Lausanne&Microsoft\\\hline
 University of Extremadura&1 & Spain& Extremadura&University of Castilla-La Mancha, University of Pisa, TU Wien, University of Cologne, University of Victoria\\\hline
 University of Notre Dame&1 & USA& Indiana&Indiana University\\\hline
 Indiana University&1 & USA& Indiana &University of Notre Dame\\\hline
 University of Cambridge & 2& United Kingdom& Cambridge&University of Delaware Newark, University of Oxford\\\hline
 University of Delaware Newark&1 & USA& Delaware&University of Cambridge, University of Oxford\\\hline
 University of Oxford& 1& United Kingdom& Oxford&University of Delaware Newark, University of Cambridge\\\hline
 Indiana University Bloomington&1 & USA& Indiana&None\\\hline
 University of Chicago&3 & USA& Chicago&University of Texas at Austin, Yale University\\\hline
 National Tsing Hua University&1 & Taiwan& Hsinchu&National Taiwan University, Columbia University, National Chiao Tung University\\\hline
 Columbia University&1 & USA& New York&National Taiwan University, National Tsing Hua University, National Chiao Tung University\\\hline
 National Chiao Tung University&1 & Taiwan& Hsinchu&National Taiwan University, National Tsing Hua University, Columbia University\\\hline
 Sorbonne Universite&1 & France& Paris&ParTec AG, Julich Supercomputing Centre\\\hline
 University of Illinois Urbana-Champaign&1 & USA& Illinois& University of British Columbia, Quantum Science \& Exploratory Research, University of Naples, Los Alamos National Laboratory,  University of Jyväskylä\\\hline
 Quantum Science \& Exploratory Research& 1& UK& Reading&University of British Columbia,  University of Naples, Los Alamos National Laboratory,  University of Jyväskylä, University of Illinois Urbana-Champaign\\\hline
 University of Naples&1 & Italy& Naples&University of British Columbia, Quantum Science \& Exploratory Research,  Los Alamos National Laboratory,  University of Jyväskylä, University of Illinois Urbana-Champaign\\\hline
 University of Jyväskylä& 2& Finland& Jyväskylä&University of British Columbia, Quantum Science \& Exploratory Research, University of Naples, Los Alamos National Laboratory,   University of Illinois Urbana-Champaign\\\hline
 Georgia Institute of Technology& 1& USA& Georgia&Google\\\hline
 King's College London&1 & UK& London&Simula Research Laboratory\\\hline
 Simula Research Laboratory&1 & Norway& Oslo&King's College London\\\hline

\end{tabular}
\end{table}

\begin{table}[htbp!]
\centering
\caption{Engagement from academia - Part IV}
\label{tab:academia4}
\begin{tabular}{|p{3cm}|l|p{2cm}|p{2.5cm}|p{8cm}|}
\hline
\textbf{Univ./research centers}&\textbf{Count}& \textbf{Country} & \textbf{Province / State} & \textbf{Collaborators }\\ \hline
 Technical University of
Applied Science Regensburg&2 & Germany& Regensburg&Siemens AG, Technology\\\hline
 City University of New York&1 & USA& New York&University of Central Florida\\\hline
 University of Central Florida&3 & USA& Florida&City University of New York, Lawrence Berkeley National Laboratory,  Coherent Computing Inc., Purdue University, Pennsylvania State University, Quantum Circuits Inc.\\\hline
 University of Catania& 1& Italy& Sicily&None\\\hline
 University of Michigan& 1& USA& Michigan&None\\\hline
 Osaka University& 1& Japan&Osaka &RIKEN Center for Quantum Computing (RQC), Fujitsu LTD, Systems Engineering Consultants Co.,LTD,  TIS Inc.\\\hline
 University College London& 1 & UK &London &Keio University, Kyoto University, RIKEN, NTT Corporation\\\hline
 Kyoto University& 1&Japan & Kyoto&Keio University,  University College London, RIKEN, NTT Corporation\\\hline
 University of Alabama in Huntsville&1 & Alabama&Huntsville &None\\\hline
 IIT Hyderabad&1 & India& Telangana& University of Maryland College Park, University of Maryland Baltimore County\\\hline
 University of Maryland Baltimore County& 1& USA&Maryland &University of Maryland College Park, IIT Hyderabad\\\hline
 University of Maryland College Park&2 & USA&Maryland &University of Maryland Baltimore County, IIT Hyderabad, RIACS at NASA Ames Research Center, University of Pittsburgh, Syracuse University,  Pacific Northwest National Laboratory, Massachusetts Institute of Technology, King’s College London, University of Pennsylvania, ETH Zurich\\\hline
 Northwestern University&1 & USA& Illinois&Yale University\\\hline
 Delft University of Technology& 2& Netherlands&Delft &Kavli Institute of Nanoscience \\\hline
 University of Castilla-La Mancha&1 & Spain& Albacete, Ciudad Real, Cuenca, Toledo, Almadén and Talavera de la Reina&University of Castilla-La Mancha, University of Pisa, TU Wien, University of Cologne, University of Victoria, University of Extremadura\\\hline
 University of Pisa& 1&Italy & Pisa&University of Castilla-La Mancha, TU Wien, University of Cologne, University of Victoria, University of Extremadura\\\hline
 TU Wien&1 &Austria & Vienna&University of Castilla-La Mancha, University of Pisa,  University of Cologne, University of Victoria, University of Extremadura \\\hline
 University of Cologne& 1&Germany &Cologne &University of Castilla-La Mancha, University of Pisa, TU Wien, University of Victoria, University of Extremadura\\\hline
University of Victoria&1 &Canada & British Columbia &University of Castilla-La Mancha, University of Pisa, TU Wien, University of Cologne, University of Extremadura\\\hline
 NC State University& 1& USA& North Carolina&None\\\hline
 Rensselaer Polytechnic Institute&1 &USA &New York &University of Texas at Dallas\\\hline
 University of Texas at Dallas&1 &USA &Texas &Rensselaer Polytechnic Institute\\\hline
 Nagoya University&1 & Japan& Nagoya&Kyushu Unversity\\\hline

\end{tabular}
\end{table}

\begin{table}[htbp!]
\centering
\caption{Engagement from academia - Part V}
\label{tab:academia5}
\begin{tabular}{|p{3cm}|l|p{2cm}|p{2.5cm}|p{8cm}|}
\hline
\textbf{Univ./research centers}&\textbf{Count}& \textbf{Country} & \textbf{Province / State} & \textbf{Collaborators }\\ \hline
 Kavli Institute of Nanoscience& & Netherlands& Delft&Delft University of Technology\\\hline
 University of Pennsylvania& 1& USA&Pennsylvania & RIACS at NASA Ames Research Center, University of Pittsburgh, Syracuse University, ETH Zurich, Pacific Northwest National Laboratory, University of Maryland, College Park, Massachusetts Institute of Technology, King’s College London, University of Pennsylvania\\\hline
 University of Pittsburgh&1 & USA& Pennsylvania&RIACS at NASA Ames Research Center,  Syracuse University, ETH Zurich, Pacific Northwest National Laboratory, University of Maryland, College Park, Massachusetts Institute of Technology, King’s College London\\\hline
 Syracuse University& 1&USA &New York & RIACS at NASA Ames Research Center, University of Pittsburgh, ETH Zurich, Pacific Northwest National Laboratory, University of Maryland, College Park, Massachusetts Institute of Technology, King’s College London, University of Pennsylvania\\\hline
 King’s College London& 1&UK &London &RIACS at NASA Ames Research Center, University of Pittsburgh, Syracuse University, ETH Zurich, Pacific Northwest National Laboratory, University of Maryland, College Park, Massachusetts Institute of Technology, University of Pennsylvania\\\hline
 ETH Zurich&1 & Switzerland &Zurich &RIACS at NASA Ames Research Center, University of Pittsburgh, Syracuse University,  Pacific Northwest National Laboratory, University of Maryland, College Park, Massachusetts Institute of Technology, King’s College London, University of Pennsylvania\\\hline
 Massachusetts Institute of Technology& 1& USA& Massachusetts &RIACS at NASA Ames Research Center, University of Pittsburgh, Syracuse University, ETH Zurich, Pacific Northwest National Laboratory, University of Maryland, College Park,  King’s College London, University of Pennsylvania \\\hline
 University of Würzburg&1 & Germany& Würzburg&None\\\hline
 University of Colorado Colorado Springs& 2& USA& Colorado&None\\\hline
 Washington University& 1&USA &Washington &Cisco, Hamad Bin Khalifa University (HBKU)\\\hline
 Hamad Bin Khalifa University (HBKU)& &Qatar &Doha &Washington University, Cisco\\\hline
 Seoul National University& 2& South Korea& Seoul&Singapore Management University, Pusan National University \\\hline
 Singapore Management University& 2& Singapore& - &Seoul National University, Pusan National University\\\hline
 Pusan National University&1 &South Korea &Busan &Seoul National University, Singapore Management University\\\hline

\end{tabular}
\end{table}

\begin{table}[htbp!]
\centering
\caption{Engagement from industry (industrial research centers and companies)}
\label{tab:industry}
\begin{tabular}{|p{4cm}|l|p{2cm}|p{2.5cm}|p{4cm}|}
\hline
\textbf{Indust. res. center/company}  &\textbf{Count}& \textbf{Country} & \textbf{Province / State} & \textbf{Collaborators }\\ \hline
Deloitte Consulting LLP  &1& United States & --- & Arizona State University (School of Computing and Augmented Intelligence) \\ \hline
Fraunhofer Institute for Manufacturing Engineering and Automation (Fraunhofer IPA)  &1& Germany & Baden--Württemberg (Stuttgart) & None (authors solely affiliated with Fraunhofer IPA) \\ \hline
 NVIDIA& 2& United States& California&Argonne National Laboratory, Lawrence Berkeley National Laboratory - Google Research (USA); Oak Ridge National Laboratory (USA); Univ. of Wisconsin–Madison (USA)\\\hline
 Fraunhofer Institute for Transportation and Infrastructure Systems (Fraunhofer IVI), Ingolstadt site& 1& Germany& Bavaria (Bayern)&None (all authors affiliated with Fraunhofer IVI).\\\hline
 Data:Lab, Volkswagen AG & 1& Germany& Bavaria (Munich)&University of Innsbruck; ParityQC (Innsbruck)\\\hline
 ParityQC (Innsbruck)& 1& Austria& Tyrol&Volkswagen AG Data:Lab (Munich); University of Innsbruck\\\hline
 IBM Quantum, IBM T.J. Watson Research Center (Yorktown Heights)& 1& United States& New York&Georgetown University; North Carolina State University; IBM Research Almaden (San Jose)\\\hline
 IBM Research Almaden / IBM Quantum (San Jose)& 1& United States& California&Georgetown University; North Carolina State University; IBM T.J. Watson Research Center\\\hline
 IBM Quantum, MIT-IBM Watson AI Lab (Cambridge)& 1& United States& Massachusetts&MIT Center for Theoretical Physics\\\hline
 Hitachi, Ltd., Research \& Development Group (Central Research Laboratory)& 1& Japan& Tokyo, Yokohama&None --- all authors  are from Hitachi R\&D.\\\hline
 Munich Quantum Software Company GmbH& 1& Germany& Bavaria&Technical University of Munich (Germany); Yale University (USA); Software Competence Center Hagenberg (Austria)\\\hline
Amazon Web Services (Amazon Braket)& 1& United States& Washington&Stanford University (USA)\\\hline
Nippon Telegraph and Telephone (NTT)& 1& Japan& Tokyo Metropolis&RIKEN Center for Quantum Computing (Japan),  Keio University, Kyoto University,  University College London\\\hline
Google Research& 1& United States& California&NVIDIA (USA); Oak Ridge National Laboratory (USA); Univ. of Wisconsin–Madison (USA)\\\hline
Hartree Centre (STFC)& 1& United Kingdom& Cheshire&Algorithmiq (Finland); Oak Ridge \& PNNL (USA); multiple universities\\\hline
Algorithmiq& 1& Finland& Uusimaa&Hartree Centre (UK); Oak Ridge \& PNNL (USA)\\\hline
Siemens AG (Foundational Technologies)& 1& Germany& Bavaria&OTH Regensburg (Germany)\\\hline
IBM Research (IBM Quantum)& 1& United States& New York&Unitary Fund (USA); academic collaborators\\\hline
Unitary Fund / Unitary Labs& 1& United States& California&IBM Research (USA); academic collaborators\\
 & & & &\\\hline
\end{tabular}
\end{table}

\end{document}